\RequirePackage{fix-cm}
\documentclass[twocolumn,epjc3]{svjour3}  
\smartqed  
\RequirePackage{graphicx}
\journalname{Eur. Phys. J. C}
\usepackage{cite} 
\usepackage{amsmath}
\usepackage{hyperref}
\usepackage{cite}
\usepackage{amsmath,amssymb}
\usepackage{color}

\begin{document}
\title{Regularity condition on the anisotropy induced by gravitational decoupling in the framework of MGD}


\author{
	G. Abell\'an \thanksref{addr2}
	\and
    V.A. Torres--S\'anchez   
	 \thanksref{addr1}        
        \and
E. Fuenmayor
     \thanksref{addr2}
\and
	 E. Contreras
\thanksref{e3,
	addr3}
}
\thankstext{e3}{e-mail: 
\href{mailto:ernesto.contreras@gmail.com}{\nolinkurl{econtreras@usfq.edu.ec}} 
}
\institute{
Grupo de Campos y Part\'iculas, Facultad de Ciencias, Universidad Central de Venezuela, AP 47270, Caracas 1041-A, Venezuela\label{addr2}
\and
 School of Physical Sciences \& Nanotechnology, Yachay Tech University, 100119 Urcuqu\'i, Ecuador \label{addr1}
\and
 Departamento de F\'isica, Colegio de Ciencias e Ingenier\'ia, Universidad San Francisco de Quito, Quito, Ecuador.\label{addr3}
}
\date{Received: date / Accepted: date}

\maketitle
\begin{abstract}
We use gravitational decoupling to establish a connection between the minimal geometric deformation approach and the standard method for obtaining anisotropic fluid solutions. Motivated by the relations that appear in the framework of minimal geometric deformation, we give an anisotropy factor that allows us to solve the quasi--Einstein equations associated to the decoupling sector. We illustrate this by building an anisotropic extension of the well known Tolman IV solution, providing in this way an exact and physically acceptable solution that represents the behavior of compact objects. We show that, in this way, it is not necessary to use the usual mimic constraint conditions. Our solution is free from physical and geometrical singularities, as expected. We have presented the main physical characteristics of our solution both analytically and graphically and verified the viability of the solution obtained by studying the usual criteria of physical acceptability.

\end{abstract}

\maketitle

\section{Introduction}\label{intro}

In 1916, Karl Schwarzschild obtained the first interior solution of the Einstein field equations \cite{sch}. This solutions describe a self--gravitating object sustained by a perfect and incompressible fluid which is embedded in a static and spherically symmetric
vacuum space--time. Following the strategy of Schwarzschild, other interior solutions can be constructed providing suitable equations of state to close the system. However, in some cases the system obtained can not be analytically integrated and numerical models are required. Besides proposing an equation of state to relate thermodynamical quantities, we can use geometrical constraints on the functions. Indeed, following this program Tolman obtained a family of eight isotropic solutions \cite{tolman}.

For many years isotropic 
solutions were considered as well posed models to study stellar interiors. However, as was shown by Delgaty and Lake \cite{lakePF}, very few of this solutions can be considered physically acceptable (for a list of physical conditions of interior solutions see, for example, \cite{ivanov}). Nevertheless, even 
when acceptable solutions can be found
the perfect fluid model is evidently not valid when local anisotropy of pressure is assumed. Regardingly,
anisotropic models have been considered as very reasonable for describing the matter distribution under a variety of circumstances \cite{lemaitre,bowers,cosenza1,cosenza2,herrera92,bondi1,barreto,matrinez,herrera97,herrera98,bondi2,hernandez,herrera2001,aurora,
	herreraallstatic,nunez,
	contrerasAS}.
Now, as it is well known, assumption of local anisotropy in the fluid leads to the introduction of an extra unknown quantity in the system. In this sense, we need to imposse two conditions, either equations of state or geometric links between the metric variables in order to integrate the system.
For example, in Ref. \cite{bowers} Bowers and Liang besides assuming the Schwarzschild constraint on the density, they impossed that in order to avoid singularities in the Tolman--Oppenheimer--Volkoff (TOV) equation
\begin{eqnarray}\label{TOV}
p'_{r}=-(\rho+p_{r})\frac{\nu'}{2}
+2\frac{p_{\perp}-p_{r}}{r}\,,
\end{eqnarray}
the anisitropy, $p_{\perp}-p_{r}$ must satisfy, for example, the following constraint
\begin{eqnarray}\label{BL}
p_{\perp}-p_{r}= C f(p_{r},r)(\rho+p_{r})r^{n}\,.
\end{eqnarray}
In the above expression $\rho$ is the energy density of the fluid and $p_{\perp}$
and $p_{r}$ stand for the transverse and radial pressure of the fluid, respectively. The function $f$ encodes the information of the anisotropy of the system which is no necessarily a linear function of the radial pressure and $C$ a parameter which measures the anisotropy strength. Finally, the exponent is constrained to $n>1$ in order to aviod singularities in (\ref{TOV}). In the same spirit, Cosenza et. al. \cite{cosenza1} proposed a
method which allowed them to find a family of non--isotropic models from any isotropic model which depends continuously on the constant $C$. The protocol consists in to take the energy density of any perfect fluid as the density of the anisotropic system and consider the following anisotropic function, $f(p_{r},r)$
\begin{eqnarray}\label{CA}
f(p_{r},r)=\frac{\nu'}{2}r^{1-n}\,.
\end{eqnarray}
Following this procedure they 
were able to extend the Schwarzshild interior solution, Tolman IV, V and VI, and the Adler model to anisotropic domains. However, it is worth mentioning that in some cases numerical analysis were required to obtain the solution. The procedure just explained corresponds to the standard way in which problems have been solved in the presence of anisotropy.
 
Recently,
the so--called Minimal Geometric Deformation (MGD) method \cite{ovalle2008,
	ovalle2009,ovalle2010,casadio2012,ovalle2013,ovalle2013a,
	casadio2014,casadio2015,ovalle2015,casadio2015b,
	ovalle2016, cavalcanti2016,casadio2016a,ovalle2017,
	rocha2017a,rocha2017b,casadio2017a,ovalle2018,ovalle2018bis,
	estrada2018,ovalle2018a,lasheras2018,gabbanelli2018,sharif2018,sharif2018a,sharif2108b,
	fernandez2018,fernandez2018b,
	contreras2018, estrada, contreras2018a,morales,tello18,
	rincon2018,ovalleplb,contreras2018c,contreras2019,contreras2019a,tello2019,	contrerasextended,tello2019a,lh2019,estrada2019,gabbanelli2019,ovalle2019a,sudipta2019,linares2019,
	leon2019,tello2019c}
has emerged as an alternative to extend isotropic solutions in a straightforward and analytical way 
given the number of ingredients which convert it in a versatile and powerful tool to solve the Einstein's equations.

For example, the method has been used to obtain anisotropic like--Tolman IV solutions \cite{ovalle2017,ovalle2018}, 
anisotropic Tolman VII solutions \cite{sudipta2019} and a model for neutron stars \cite{victorNS}. In other contexts, MGD has been used to
extend black holes in $3+1$ and $2+1$ dimensional space--times \cite{ovalle2018a,contreras2018a,contreras2018c}. Moreover, in the context of modified theories of gravitation, the method has been used to obtain solutions in $f(\mathcal{G})$ gravity \cite{sharif2018a}, Lovelock \cite{estrada2019}, $f(R,T)$ \cite{tello2019a} and more recently interior solutions in the context of braneworld \cite{leon2019}.

It is worth mentioning that in contrast to the standard strategy followed in Ref. \cite{cosenza1} where the information of the isotropic solution entered via the energy density of a well known model, in
the MGD method the isotropic solution is a sector of the total solution. More precisely, the isotropic solution is used as a seed to obtain anisotropic solutions of the Einstein equations as follows.

Let us consider the Einstein field equations
\begin{eqnarray}
R_{\mu\nu}-\frac{1}{2}R g_{\mu\nu}=-\kappa^{2}T_{\mu\nu}^{(tot)}\,,
\end{eqnarray}
and assume that the total energy--momentum tensor, $T_{\mu\nu}^{(tot)}$, can be decomposed as
\begin{eqnarray}\label{total}
T_{\mu\nu}^{(tot)}=T_{\mu\nu}^{(m)}+\alpha\theta_{\mu\nu}\;,
\end{eqnarray}
where $T^{(m)}_{\mu\nu}$ is the matter energy momentum for a perfect fluid and $\theta_{\mu\nu}$ an anisotropic source interacting with $T^{(m)}_{\mu\nu}$. Note that,
since the Einstein tensor is divergence free, the total energy momentum tensor $T_{\mu\nu}^{(tot)}$
satisfies
\begin{eqnarray}\label{cons}
\nabla_{\mu}T^{(tot)\mu\nu}=0.
\end{eqnarray}

It is important to point out that, as this equation is fulfilled and given that for a perfect fluid we also have
$\nabla_\mu T^{(m)\mu\nu} = 0$, then the following condition necessarily must be satisfied
\begin{equation}\label{mgd10}
    \nabla_\mu \theta^{\mu\nu} = 0\;.
\end{equation}
In this
sense, there is no exchange of energy--momentum tensor between the perfect fluid and the
anisotropic  source and henceforth interaction is purely gravitational.

In what follows, we shall consider a static, spherically symmetric space--time with line element
parameterized as
\begin{eqnarray}\label{le}
ds^{2}=e^{\nu}dt^{2}-e^{\lambda}dr^{2}-r^{2}d\Omega^{2}\,,
\end{eqnarray}
where $\nu$ and $\lambda$ are functions of the radial coordinate $r$ only. 
Now, considering Eq. (\ref{le}) as a solution of the Einstein equations, we obtain
\begin{eqnarray}
\kappa^{2} \tilde{\rho}&=&\frac{1}{r^{2}}+e^{-\lambda}\left(\frac{\lambda'}{r}-\frac{1}{r^{2}}\right),\label{eins1}\\
\kappa^{2} \tilde{p}_{r}&=&-\frac{1}{r^{2}}+e^{-\lambda}\left(\frac{\nu'}{r}+\frac{1}{r^{2}}\right),\label{eins2}\\
\kappa^{2} \tilde{p}_{\perp}&=&\frac{e^{-\lambda}}{4}\left(\nu'^{2}-\nu '\lambda '+2\nu''
+2\frac{\nu'-\lambda'}{r}\right)\label{eins3},
\end{eqnarray}
where the primes denote derivation with respect to the radial coordinate and we have defined
\begin{eqnarray}
\tilde{\rho}&=&\rho+\alpha\theta^{0}_{0}\,,\label{rot}\\
\tilde{p}_{r}&=&p-\alpha\theta^{1}_{1}\,,\label{prt}\\
\tilde{p}_{\perp}&=&p-\alpha\theta^{2}_{2}\,.\label{ppt}
\end{eqnarray} 
Note that, at this point, the decomposition (\ref{total}) seems as a simple separation of the constituents of the matter sector. Even more, given the non--linearity of Einstein's equations, such a decomposition does not lead to a decoupling of two set of equations, one for each source involved. However, contrary to the broadly belief, the decoupling is possible in the context of MGD. The method consists in to introduce a geometric deformation in the metric functions given by
\begin{eqnarray}
\nu&=&\xi+\alpha g\,,\\
e^{-\lambda}&=&\mu +\alpha f\label{def}\,,
\end{eqnarray}
where $\{g,f\}$ are the so--called decoupling functions and $\alpha$ is a free parameter that
``controls'' the deformation. It is worth mentioning that although a general treatment considering deformation in both components of the metric is possible (see Ref. \cite{ovalleplb}), in this work we shall concentrate in the particular case $g=0$ and $f\ne 0$.  Doing so, we obtain
two sets of differential equations: one describing an isotropic system sourced by
the conserved energy--momentum tensor of a perfect fluid $T^{(m)}_{\mu\nu}$ and the other
set corresponding to quasi--Einstein field equations sourced by $\theta_{\mu\nu}$. More precisely, 
we obtain
\begin{eqnarray}
\kappa ^2 \rho &=&\frac{1-r \mu'-\mu}{r^{2}}\,,\label{iso1}\\
\kappa ^2 p&=&\frac{r \mu  \nu '+\mu -1}{r^{2}}\,,\label{iso2}\\
\kappa ^2 p&=&\frac{\mu ' \left(r \nu '+2\right)+\mu  \left(2 r \nu ''
+r \nu '^2+2 \nu '\right)}{4 r}\label{iso3}\,,
\end{eqnarray}
with
\begin{eqnarray}
\nabla_{\mu}T^{(m)\mu\nu}=p'
-\frac{\nu'}{2}(\rho+p)=0\,,
\end{eqnarray}
for the perfect fluid
and
\begin{eqnarray}
\kappa ^2  \theta^{0}_{0}&=&-\frac{r f'+f}{r^{2}}\,,\label{aniso1}\\
\kappa ^2 \theta^{1}_{1}&=&-\frac{r f \nu '+f}{r^{2}}\,,\label{aniso2}\\
\kappa ^2\theta^{2}_{2}&=&-\frac{f' \left(r \nu '+2\right)+f \left(2 r \nu ''+r \nu '^2+2 \nu '\right)}{4 r}\label{aniso3}\,,
\end{eqnarray}
for the source $\theta_{\mu\nu}$ that, whenever 
$\theta^{1}_{1}\ne\theta^{2}_{2}$, induce local anisotropy in the system as can be seen in Eqs. (\ref{prt}) and (\ref{ppt}). It is worth noticing that the conservation equation $\nabla_{\mu}\theta^{\mu}_{\nu}=0$ leads to
\begin{eqnarray}\label{tovtheta}
(\theta^{1}_{1})'-\frac{\nu'}{2}(\theta^{0}_{0}-\theta^{1}_{1})-\frac{2}{r}(\theta^{2}_{2}-\theta^{1}_{1})=0\,.
\end{eqnarray}
which is a linear combination of Eqs. (\ref{aniso1}), (\ref{aniso2}) and (\ref{aniso3}). Note that unlike quasi--Einstein equations, which differ from the Einstein equations, this equation is completely analogous to an anisotropic TOV equation as can be seen in Ref. \cite{cosenza1}.

Now,  given  metric functions $\{\nu,\mu\}$ 
sourced by a perfect fluid $\{\rho,p\}$ that solve Eqs. (\ref{iso1}), (\ref{iso2}) and (\ref{iso3}),
the deformation function $f$ can be found from Eqs. (\ref{aniso1}), (\ref{aniso2}) and (\ref{aniso3})
after choosing suitable conditions on the anisotropic source $\theta_{\mu\nu}$. It is worth mentioning
that the case we are dealing with demands for an exterior Schwarzschild solution. In this case, the
matching condition leads to the extra information required to completely solve the system. 

Defining $\mu(r)=1-\frac{2m(r)}{r}$ in (\ref{def}), the interior solution parameterized with
(\ref{le}) reads
\begin{eqnarray}
ds^{2}=e^{\nu}\!dt^{2}-\left(1-\frac{2m(r)}{r}+\alpha f\right)^{-1}\!\!dr^{2}-r^{2}d\Omega^{2}\,.
\end{eqnarray}
Now, outside of the distribution the space--time is that of Schwarzschild, given by
\begin{eqnarray}
ds^{2}=\left(1-\frac{2M}{r}\right)\!dt^{2}-\left(1-\frac{2M}{r}\right)^{-1}\!\!dr^{2}-r^{2}d\Omega^{2}.
\end{eqnarray}
In order to match smoothly the two metrics above on the
boundary surface $\Sigma$, we must require the continuity of
the first and the second fundamental form across that
surface. Then it follows
\begin{eqnarray}
e^{\nu_{\Sigma}}=1-\frac{2 M}{r_{\Sigma}}\,, \label{coupling01} \\
e^{-\lambda_{\Sigma}}=1-\frac{2 M}{r_{\Sigma}}\,, \label{coupling02} \\
\tilde{p}_{r_{\Sigma}}=0\,. \label{coupling03}
\end{eqnarray}
Note that, the condition on the radial pressure leads to
\begin{eqnarray}\label{cond}
p(r_{\Sigma})-\alpha\theta^{1}_{1}(r_{\Sigma})=0\,.
\end{eqnarray}
Regardingly, if the original perfect fluid match smoothly with the Schwarzschild solution, i.e, 
$p(r_{\Sigma})=0$, Eq. (\ref{cond}) can be satisfied by demanding $\theta^{1}_{1}\sim p$. Of course,
the simpler way to satisfy the requirement on the radial pressure is assuming the so--called mimic constraint \cite{ovalle2017} for the pressure, namely
\begin{eqnarray}\label{mimetic}
\theta^{1}_{1}= p\,,
\end{eqnarray}
in the interior of the star. 
	Remarkably, this condition leads to an algebraic equation for $f$ such that, in principle, any isotropic solution can be extended with this constraint. Another possibility is to use the mimic constraint for the density which leads to a differental equation for $f$ which can be solved in some situations (see for example \cite{ovalle2018}).
However, as far as we know, no physical requirements on the anisotropy function induced by the decoupling sector, $\theta^{2}_{2}-\theta^{1}_{2}$, have been considered up to now. In this work we find an anisotropic solution assuming a regularity condition on the anysotropy function of the decoupling sector following Bowers--Liang constraint, given by Eq. (\ref{BL}) and the Cosenza--Herrera--Esculpi--Witten  anisotropy defined in Eq. (\ref{CA}). In this sense, we propose the following condition on the  decoupling sector reads,
\begin{eqnarray}\label{BLN}
\theta^{2}_{2}-\theta^{1}_{1} = C f(\theta^{1}_{1},r)(-\theta^{0}_{0}+\theta^{1}_{1})r^{n}\,,
\end{eqnarray}
with $f(\theta^{1}_{1},r)r^{n-1}=\nu'/2$ and $C$, as usual, is a constant that gauge the anisotropy strength. This ansatz is inspired by the relation between the components of the anisotropic energy--momentum tensor $\theta^{\mu\nu}$ and the effective quantities given by Eqs. (\ref{rot}), (\ref{prt}) and (\ref{ppt}). Note that the function $f$ in Eq. (\ref{BLN}) is not the deformation function that appears in Eq. (\ref{def}).
It is clear that the replacement of (\ref{aniso1}), (\ref{aniso2}) and (\ref{aniso3}) in (\ref{BLN}), leads to a differential equation for the deformation function, $f$,  where the only required information is the metric function $\nu$, which in the context of MGD is common for the three sectors involved.

This work is organized as follows. In the next section we study the regularity condition on the decoupling sector induced by MGD. In section \ref{PC}, we study the conditions for physical viability in interior solutions. Finally, the last section is devoted to final remarks.

\section{Regularity condition on the decoupling sector}\label{new}
 In the context of MGD the anisotropy is induced by the decoupling sector sourced by $\theta_{\mu\nu}$ which satisfy a conservation equation given by
(\ref{tovtheta}). Now, from Eq. (\ref{BLN}) and after impossing the Consenza--Herrera--Esculpi--Witten anisotropy we obtain
\begin{eqnarray}
&&\left[(2 C +1)r \nu '+2\right]f'+\nonumber\\
&&\hspace{20pt} + \Big\{ \left[r(1-2 C ) \nu '-2\right]\nu '+2 r \nu ''-\frac{4}{r}\Big\}f=0\, ,
\end{eqnarray}
which can be formally solved to obtain
\begin{eqnarray}\label{diff}
f=c_1 e^{ \int \frac{u \left(\nu ' \left((2 C-1) u 
	\nu '+2\right)-2 u \nu ''\right)+4}{u \left((2 C + 1) u \nu '+2\right)} \, du}\,,
\end{eqnarray}
where $c_1$ is a constant of integration. Of course, finding an analytical solution of the above integral will depend on the particular form of the metric function $\nu$. 

As a particular case of application we shall consider the Tolman IV solution given by 
\begin{eqnarray}
e^{\nu}&=&B^2 \left(1 + \frac{r^2}{A^2}\right),\label{tol1}\\
\mu&=&\frac{\left(1 + \frac{r^2}{A^2}\right) \left(1-\frac{r^2}{d^2}\right)}{1 + \frac{2 r^2}{A^2}}\label{tol2}\,,
\end{eqnarray}
where $A$, $B$ and $d$ are constants. Next, replacing (\ref{tol1}) in (\ref{diff}) we obtain
\begin{eqnarray}
f=\frac{c_1 r^2 \left(A^2+r^2\right)}{A^2+2 (C+1) r^2}\,,
\end{eqnarray}
which determines the decoupling sector completely and enusure the regularity of the anisitropy $\theta^{2}_{2}-\theta^{1}_{1}$. To complete the MGD program, the rest of the section is devoted to obtain the total like--Tolman IV anisitropic solution. From Eq. (\ref{def}), the $g^{rr}$ component of the metric reads
\begin{eqnarray}
e^{-\lambda}=\left(A^2+r^2\right) \left[\frac{\alpha  c_{1} r^2}{A^2+2 (C+1) r^2}+\frac{d^{2}-r^{2}}{d^2 \left(A^2+2 r^2\right)}\right].\nonumber\\
\end{eqnarray}
Now, from (\ref{eins1}), (\ref{eins2}) and (\ref{eins3}) we obtain the effective quantities
\begin{eqnarray}
\tilde{\rho}&=&\frac{r^2 \left(7 A^2+2 d^2\right)+3 A^2 \left(A^2+d^2\right)+6 r^4}{8\pi d^2 \left(A^2+2 r^2\right)^2}\nonumber\\
&&-\frac{\alpha  c_{1} \left[3 A^4+A^2 (2 C + 7) r^2+6 (C + 1) r^4\right]}{8\pi\left[A^2+2 (C + 1) r^2\right]^2}\,,\\
\tilde{p}_{r} &=& \frac{d^2-A^2-3 r^2}{8\pi d^2 \left(A^2+2 r^2\right)} + \frac{\alpha  c_{1} \left(A^2+3 r^2\right)}{8\pi[A^2+2 (C+1) r^2]}\, ,\\
\tilde{p}_{\perp}&=&
\frac{d^2-A^2-3 r^2}{8\pi d^2 \left(A^2+2 r^2\right)} \nonumber\\
&& + \frac{\alpha  c_{1} \left[A^4+5 A^2 r^2+6 (C+1) r^4\right]}{8\pi\left[A^2+2 (C+1) r^2\right]^2}\,.
\end{eqnarray}
In order to match the interior solution with the Schwarz-schild exterior solution, we proceed to impose the continuity of the first and the second fundamental (see Eqs. (\ref{coupling01}), (\ref{coupling02}) and (\ref{coupling03})) from where
\begin{eqnarray}
d^2 &=& \frac{(A^2+3 R^2) (A^2+2 (C+1) R^2)}{[A^2+\alpha  c_{1} \left(A^4+5 A^2 R^2+6 R^4\right)+2 (C+1) R^2]}\,, \nonumber\\
\\
B^2 &=& \frac{R-2 M}{R+\frac{R^3}{A^2}}\,,\\
A^2 &=& \frac{R^2 (R-3 M)}{M}\,.
\end{eqnarray}

In this sense, the solution is parameterized by the mass $M$, the radius $R$, the parameter of anisotropy strength $C$, the MGD parameter $\alpha$, and the constant of integration $c_{1}$. 
In the next section we shall study the 
acceptability conditions for the like--Tolman IV anistropic solution obtained here.

\section{Conditions for physical viability of interior solutions}\label{PC} 
In this section we perform the physical analysis of the properties of the star solution by fixing the free parameters of the solution and providing plots. More precisely, in order to obtain a useful model for a compact anisotropic star we specify the mass $M$ and the radius $R$ of the star and impose some suitable conditions that the model should satisfy. The following conditions have been typically recognized as decisive for anisotropic fluid spheres.

\subsection{Matter sector}
\begin{figure}[h!]
	\centering
	\includegraphics[scale=0.52]{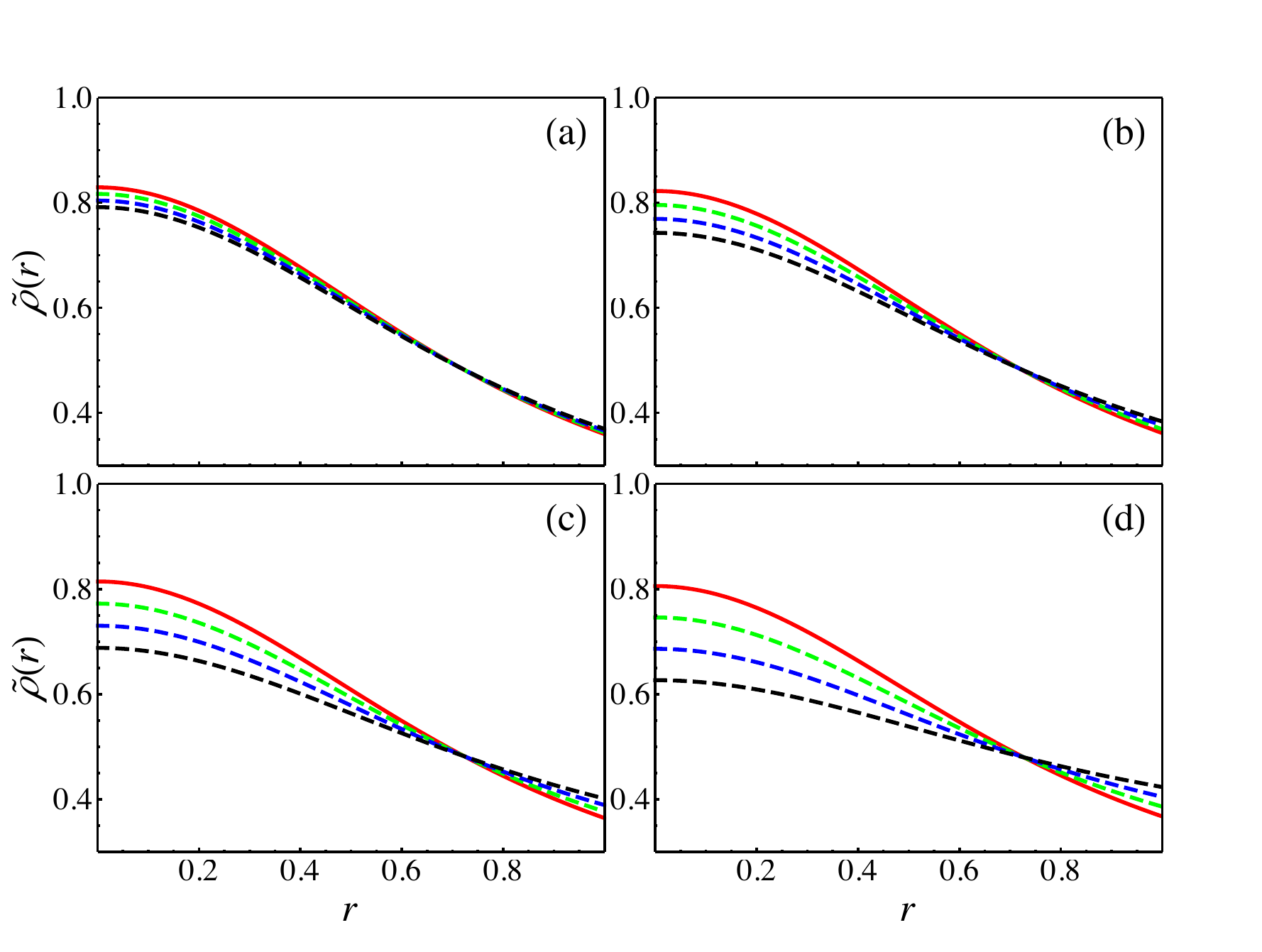}
	\caption{\label{density} 
		Energy density $\tilde{\rho}$ for $M=0.2$, $R=1$, $c_{1}=0.5$ and 
		a) $C=-0.1$, b) $C=-0.2$, c) $C=-0.3$, d) $C=-0.4$.   $\alpha=0.1$ (red line), $\alpha=0.3$, (green line),
		$\alpha=0.5$ (blue line), $\alpha=0.7$ (black line)
	}
\end{figure}

\begin{figure}[h!]
	\centering
	\includegraphics[scale=0.52]{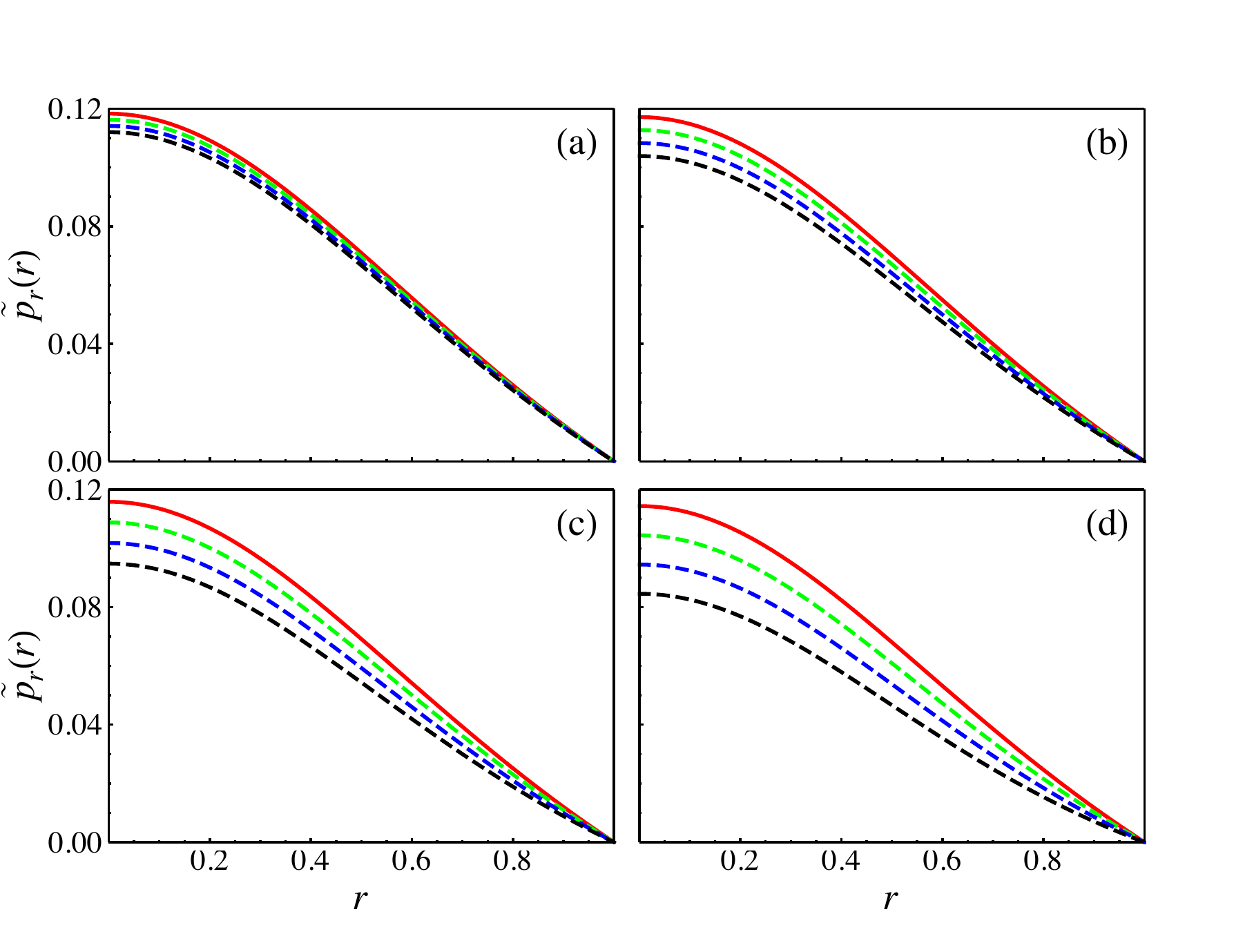}
	\caption{\label{radpress} 
		Effective radial pressure  $\tilde{p}_{r}$ for $M=0.2$, $R=1$, $c_{1}=0.5$ and 
		a) $C=-0.1$, b) $C=-0.2$, c) $C=-0.3$, d) $C=-0.4$.   $\alpha=0.1$ (red line), $\alpha=0.3$, (green line),
		$\alpha=0.5$ (blue line), $\alpha=0.7$ (black line).
	}
\end{figure}

\begin{figure}[h!]
	\centering
	\includegraphics[scale=0.52]{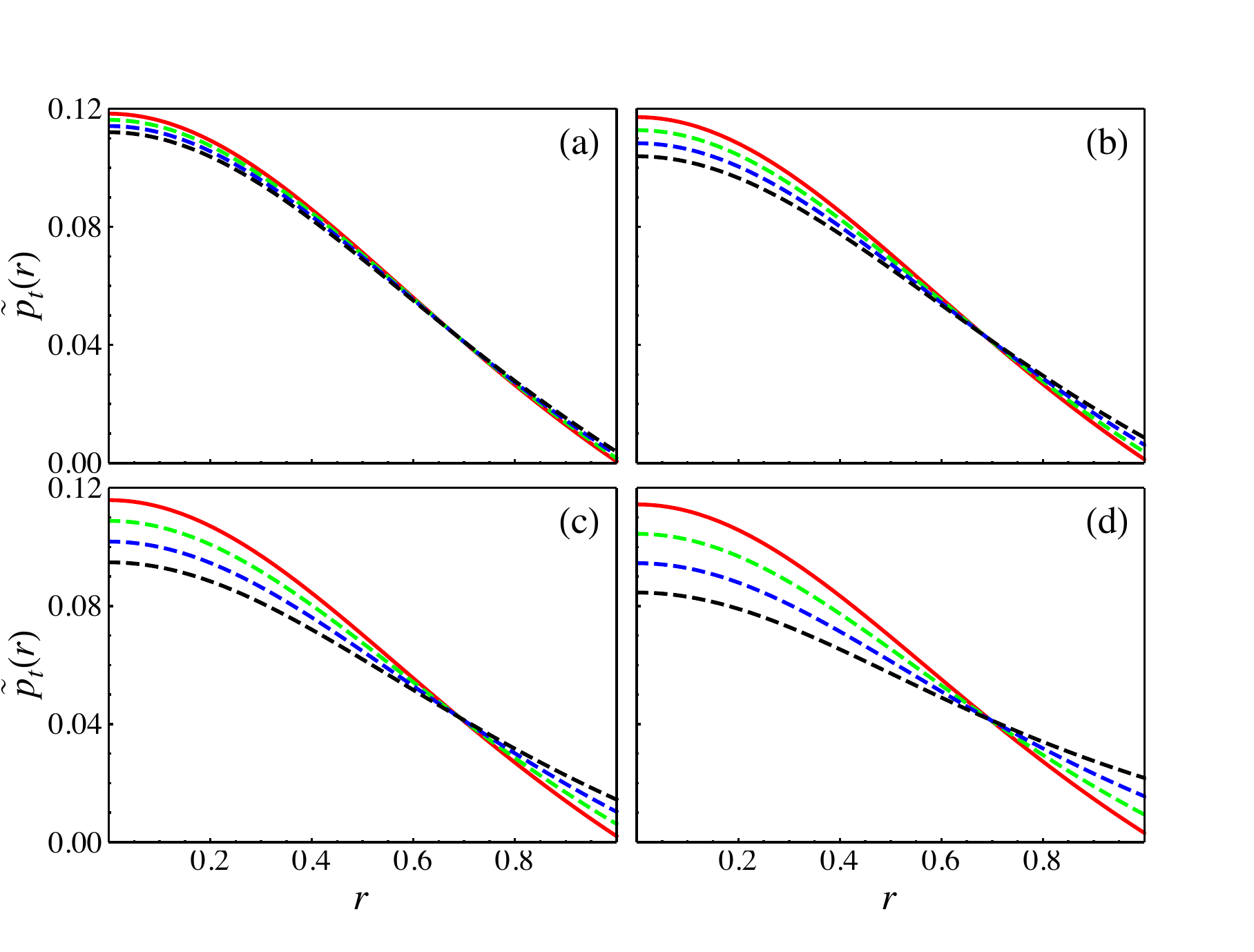}
	\caption{\label{perppress} 
		Effective tangential pressure  $\tilde{p}_{t}$ for $M=0.2$, $R=1$, $c_{1}=0.5$ and 
		a) $C=-0.1$, b) $C=-0.2$, c) $C=-0.3$, d) $C=-0.4$.   $\alpha=0.1$ (red line), $\alpha=0.3$, (green line),
		$\alpha=0.5$ (blue line), $\alpha=0.7$ (black line).
	}
\end{figure}

A requirement on the matter sector to ensure 
acceptable interior solutions is that the density and pressures should be
positive quantities. Besides, it also demanded that the density and pressures reach a maximum at the center and decrease monotonously toward the surface so that $\tilde{p}_{\perp}\ge \tilde{p}_{r}$. In figures \ref{density}, \ref{radpress} and \ref{perppress} we show the behaviour of $\tilde{\rho}$, $\tilde{p}_{r}$ and $\tilde{p}_{\perp}$  respectively.

It is worth noticing that the anisotropy factor $C$ 
has an appreciable effect on the plots in the sense that the separation of each graphic parametrized with $\alpha$ increases as $C$ decreases. To be more precise, in all the cases the  profiles in panel a) are almost indistinguishable in contrast to panel c) where the behaviour for different $\alpha$ is appreciable different.

\begin{figure}[h!]
	\centering
	\includegraphics[scale=0.52]{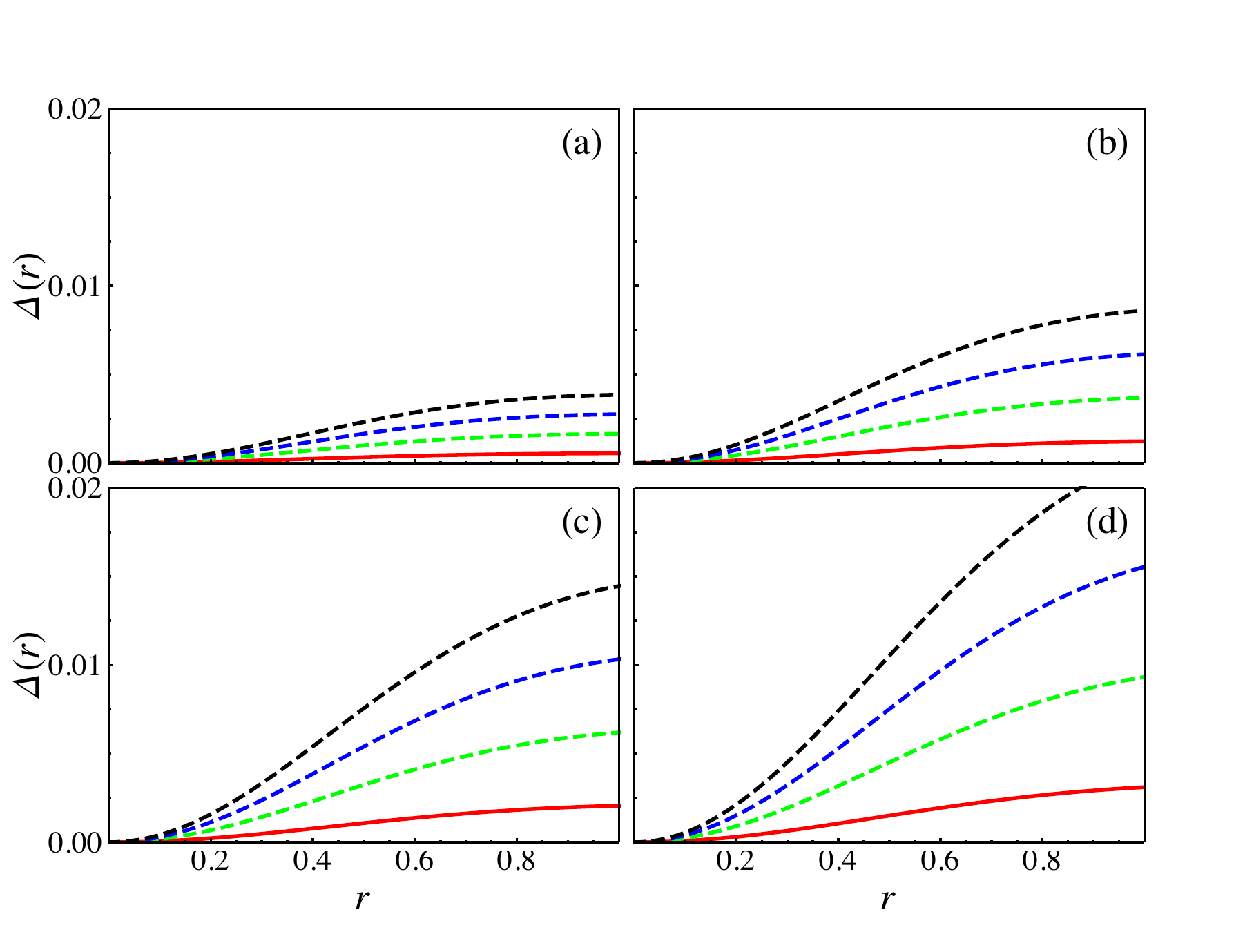}
	\caption{\label{anisotropy} 
Anisotropy  $\Delta=\tilde{p}_{t}-\tilde{p}_{r}$ for $M=0.2$, $R=1$, $c_{1}=0.5$ and 
a) $C=-0.1$, b) $C=-0.2$, c) $C=-0.3$, d) $C=-0.4$.   $\alpha=0.1$ (red line), $\alpha=0.3$, (green line),
$\alpha=0.5$ (blue line), $\alpha=0.7$ (black line).
	}
\end{figure}

To study the extra condition $\tilde{p}_{\perp}>\tilde{p}_{r}$, in fig. \ref{anisotropy} it is shown the
anisotropy $\Delta=\tilde{p_{\perp}}-\tilde{p}_{r}$. Note that the anisotropy $\Delta$ is a positive and increasing function, as expected.

\subsection{Energy conditions}
\begin{figure}[h!]
	\centering
	\includegraphics[scale=0.52]{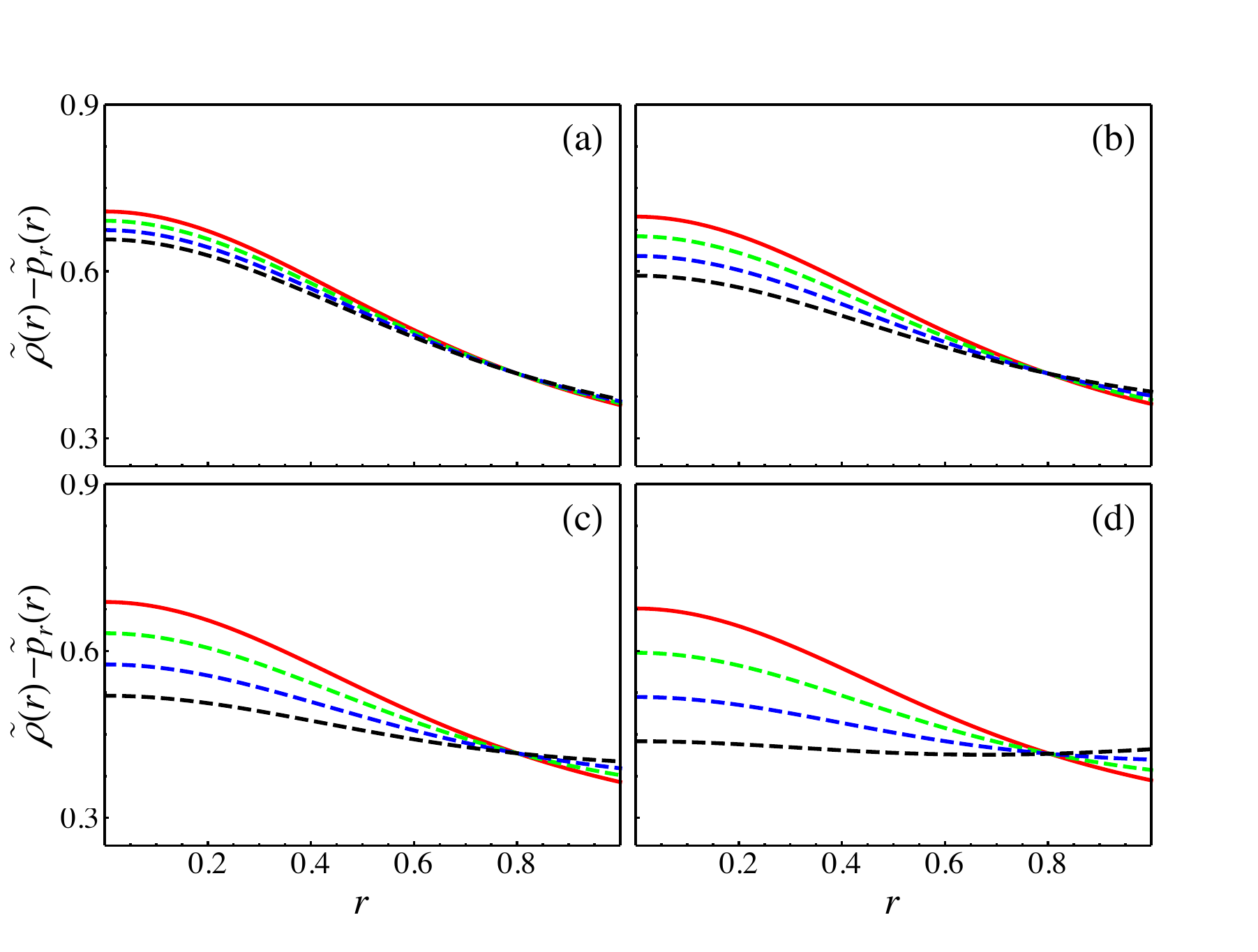}
	\caption{\label{dec1} 
DEC  for $M=0.2$, $R=1$, $c_{1}=0.5$ and 
a) $C=-0.1$, b) $C=-0.2$, c) $C=-0.3$, d) $C=-0.4$.   $\alpha=0.1$ (red line), $\alpha=0.3$, (green line),
$\alpha=0.5$ (blue line), $\alpha=0.7$ (black line).
	}
\end{figure}

Another physical requirement we demand for interior solutions involve the energy conditions. As it is well known, an acceptable interior stellar should satisfy the  dominant energy condition (DEC), which implies that  the  speed of energy flow of matter is less than the speed of light for any observer. This condition reads
\begin{eqnarray}
\tilde{\rho}-\tilde{p}_{r}&\ge& 0\,,\\
\tilde{\rho}-\tilde{p}_{\perp}&\ge& 0\,.
\end{eqnarray}
In figure \ref{dec1}, it is shown that DEC is fulfilled by all the parameters considered here.

\begin{figure}[h!]
	\centering
	\includegraphics[scale=0.52]{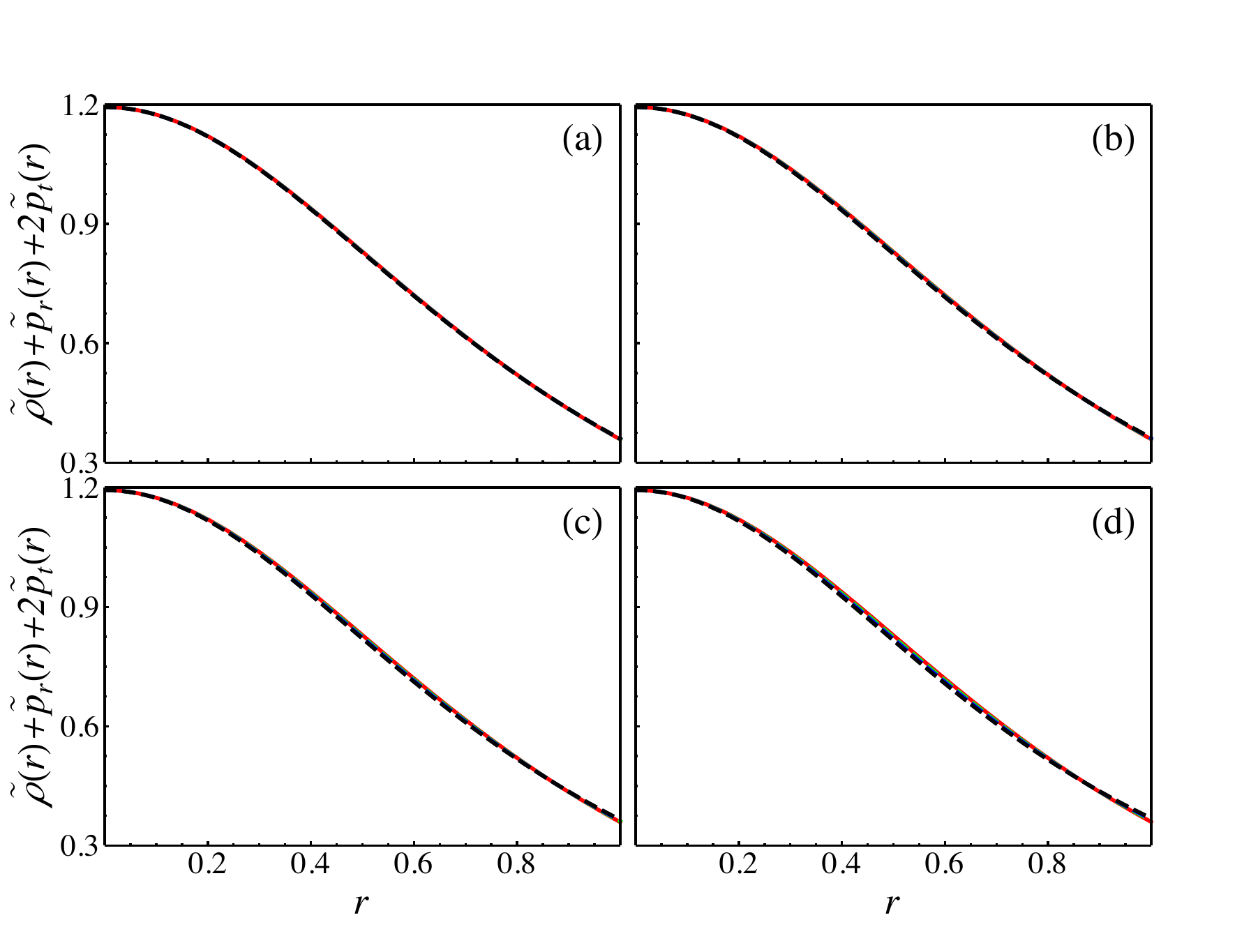}
	\caption{\label{strong} 
SEC for $M=0.2$, $R=1$, $c_{1}=0.5$ and 
a) $C=-0.1$, b) $C=-0.2$, c) $C=-0.3$, d) $C=-0.4$.   $\alpha=0.1$ (red line), $\alpha=0.3$, (green line),
$\alpha=0.5$ (blue line), $\alpha=0.7$ (black line).
	}
\end{figure}

Another condition is that the solution satisfies the strong energy condition (SEC) also, namely
\begin{eqnarray}
\tilde{\rho} + \sum\limits_{i}\tilde{p}_{i}\ge 0\,,
\end{eqnarray}
As can be seen in fig. \ref{strong}, the SEC is satisfied in the cases under consideration.

\subsection{Causality}
\begin{figure}[h!]
	\centering
	\includegraphics[scale=0.52]{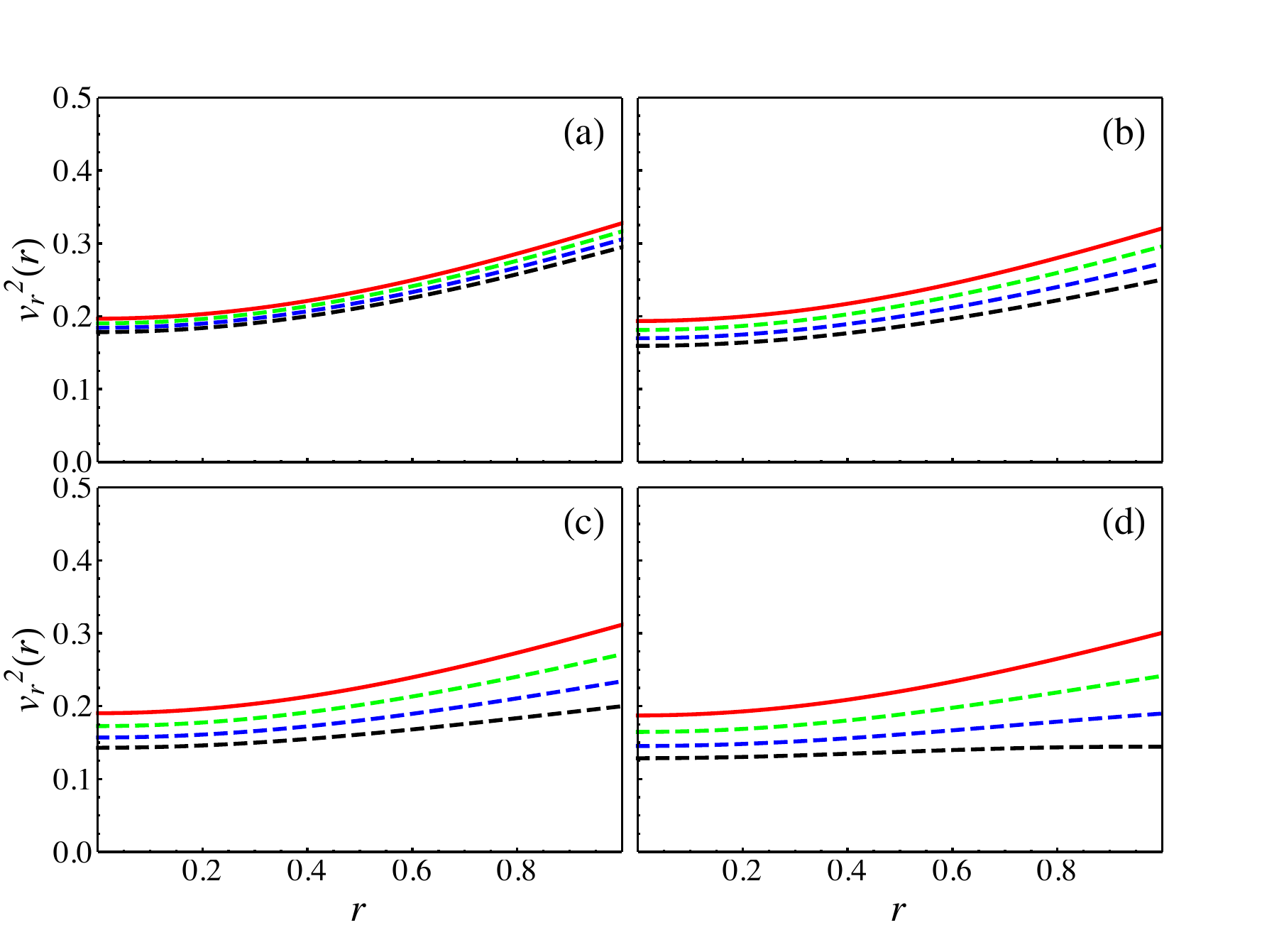}
	\caption{\label{vr} 
Radial velocity $v_{r}^{2}$ for $M=0.2$, $R=1$, $c_{1}=0.5$ and 
a) $C=-0.1$, b) $C=-0.2$, c) $C=-0.3$, d) $C=-0.4$.   $\alpha=0.1$ (red line), $\alpha=0.3$, (green line),
$\alpha=0.5$ (blue line), $\alpha=0.7$ (black line).
	}
\end{figure}

\begin{figure}[h!]
	\centering
	\includegraphics[scale=0.52]{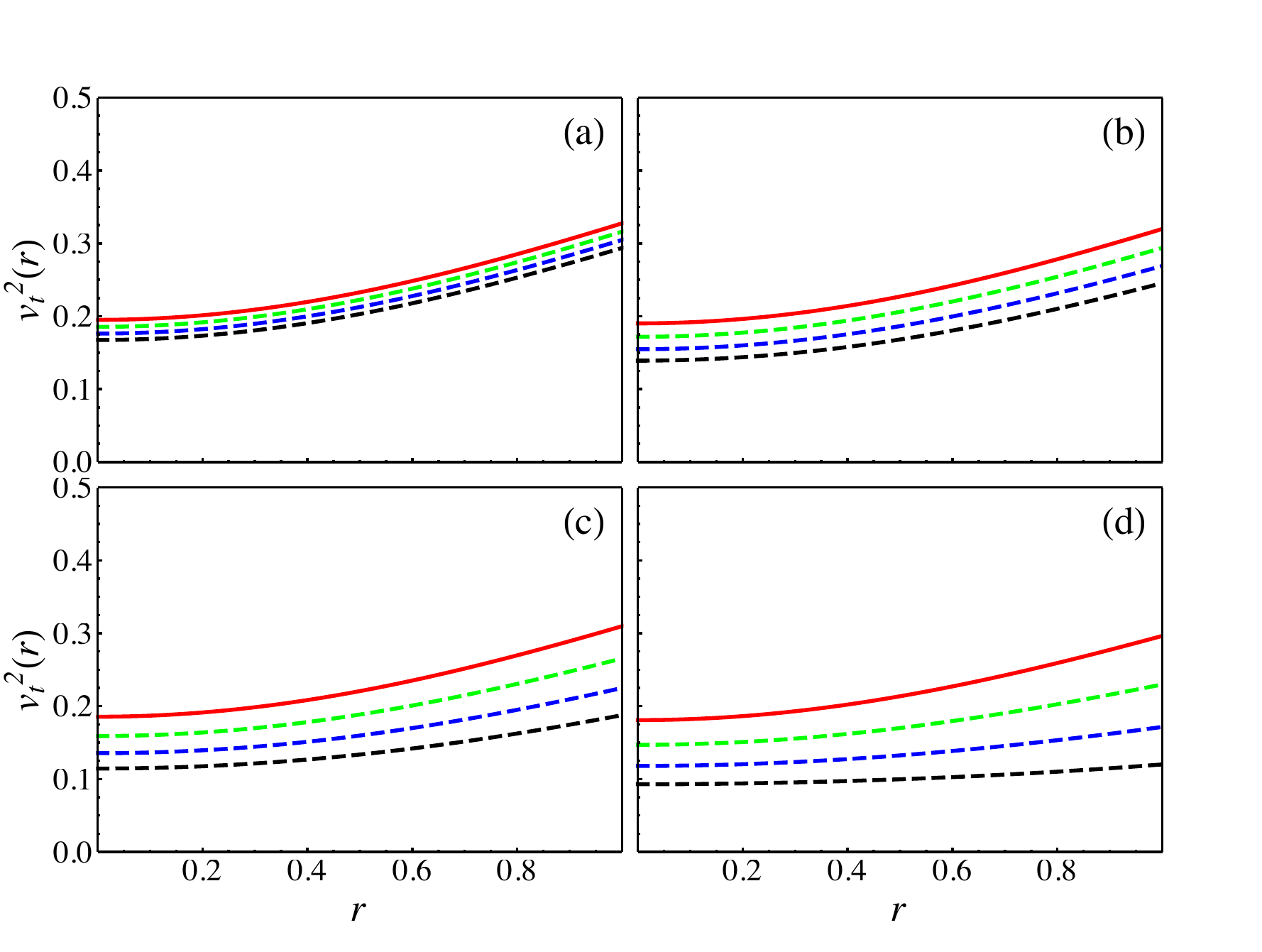}
	\caption{\label{vp} 
Tangential velocity  $v_{t}^{2}$ for $M=0.2$, $R=1$, $c_{1}=0.5$ and 
a) $C=-0.1$, b) $C=-0.2$, c) $C=-0.3$, d) $C=-0.4$.   $\alpha=0.1$ (red line), $\alpha=0.3$, (green line),
$\alpha=0.5$ (blue line), $\alpha=0.7$ (black line).
	}
\end{figure}

Causality is important to avoid superluminal motion. In other words,
the causality condition demands that
either the radial and tangential sound velocities, $v_{r}=d\tilde{p}_{r}/d\tilde{\rho}$ and $v_{t}=d\tilde{p}_{\perp}/d\tilde{\rho}$ respectively, are less than the speed of light. Given the behaviour of the radial and the traverse velocities illustrated in figures \ref{vr} and \ref{vp}, we conclude that our model satisfy the causality condition requirement.

\subsection{Adiabatic index}
\begin{figure}[h!]
	\centering
	\includegraphics[scale=0.52]{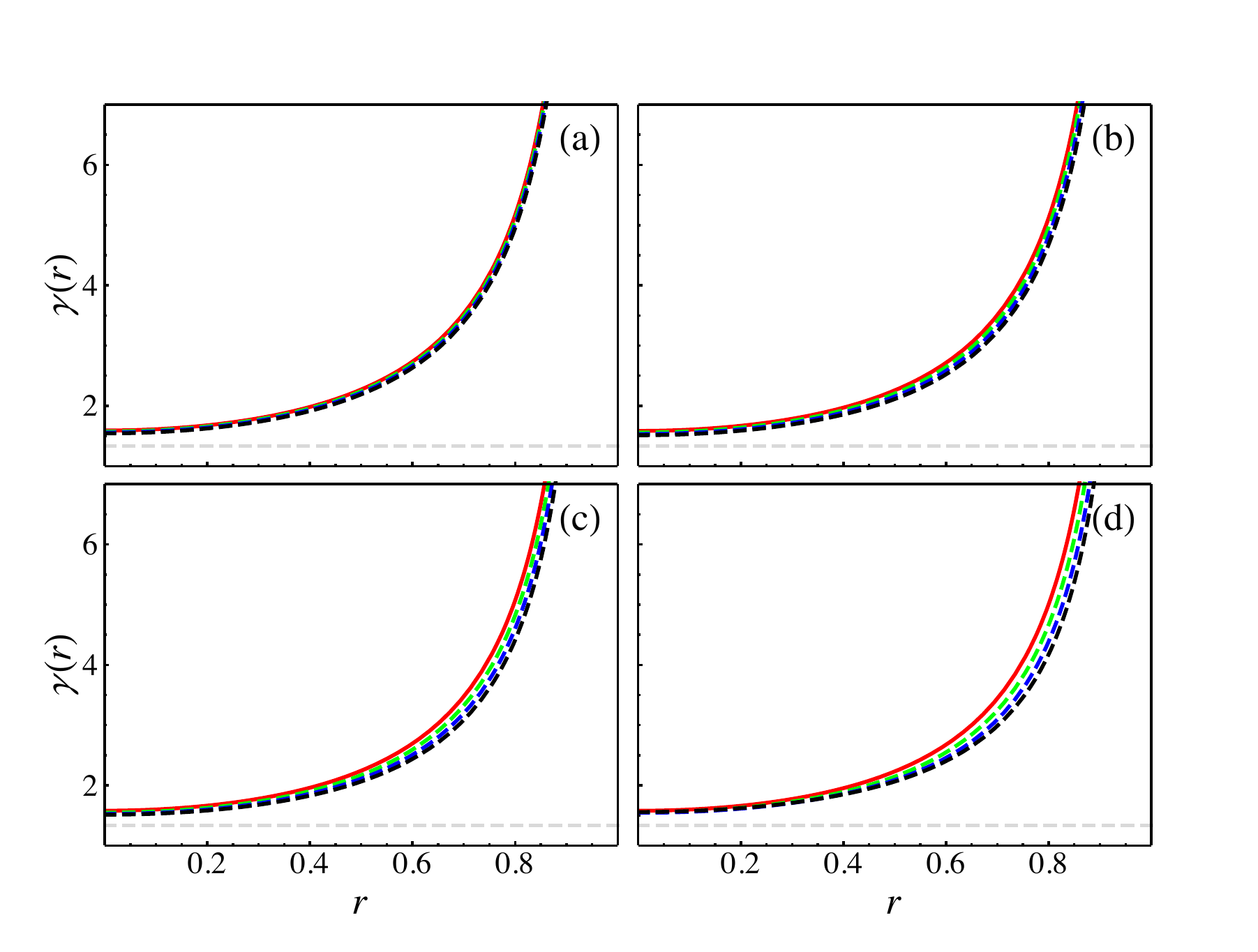}
	\caption{\label{ai} 
Adiabatic index $\gamma$ for $M=0.2$, $R=1$, $c_{1}=0.5$ and 
a) $C=-0.1$, b) $C=-0.2$, c) $C=-0.3$, d) $C=-0.4$.   $\alpha=0.1$ (red line), $\alpha=0.3$, (green line),
$\alpha=0.5$ (blue line), $\alpha=0.7$ (black line).
	}
\end{figure}

The adiabatic index, $\gamma$, serves as a criterion of stability of the interior solution.  It can be shown that for anisotropic fluids the adiabatic index takes the form
\begin{eqnarray}
\gamma=\frac{\tilde{\rho}+\tilde{p}_{r}}{\tilde{p}_{r}}\frac{d\tilde{p}_{r}}{d\tilde{\rho}}\,,
\end{eqnarray}
It is said that an interior configuration is stable whenever $\gamma\ge4/3$. In figure \ref{ai} we show the adiabatic index for different values of the free parameters involved. It is clear that the solution is stable regarding the adiabatic index criterion.


\subsection{Stability against gravitational cracking}
\begin{figure}[h!]
	\centering
	\includegraphics[scale=0.52]{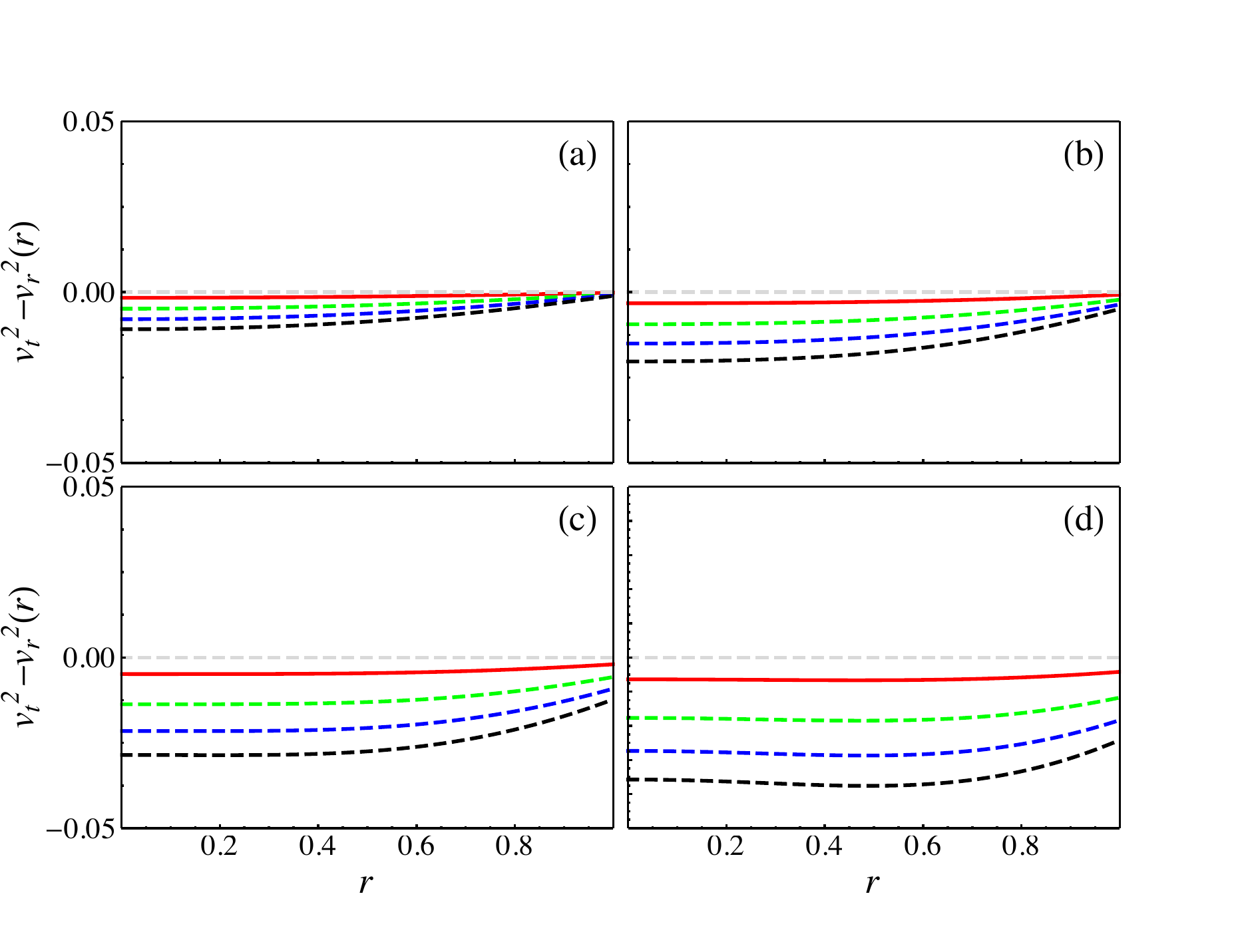}
	\caption{\label{crack} 
Anti--cracking condition   for $M=0.2$, $R=1$, $c_{1}=0.5$ and 
a) $C=-0.1$, b) $C=-0.2$, c) $C=-0.3$, d) $C=-0.4$.   $\alpha=0.1$ (red line), $\alpha=0.3$, (green line),
$\alpha=0.5$ (blue line), $\alpha=0.7$ (black line).
	}
\end{figure}

The appearance of non--vanishing total radial force with different signs in different regions of the fluid is a sign of instability. When this radial force, running from the center to the outside of the star, shifts from pointing to the center to pointing outward, the phenomenon has been called gravitational cracking \cite{lhcrack}. In reference \cite{abreu} it is stated that a simple requirement to avoid gravitational cracking is
\begin{eqnarray}\label{eqc}
-1\le\frac{d\tilde{p}_{\perp}}{d\tilde{\rho}}-
\frac{d\tilde{p}_{r}}{d\tilde{\rho}}\le 0\,.
\end{eqnarray}
In figure \ref{crack} we show that in all the cases considered the solution is stable against gravitational cracking.


\section{Final remarks}\label{remarks}
The minimal geometric deformation method has proven to be a simple and powerful tool for obtaining solutions of Einstein's field equations. The model studied in this article describing anisotropic fluid spheres meets all the requirements to be an acceptable solution.

In this work we established a connection between standard approaches to obtain anisotropic stellar solutions and the minimal geometric deformation method. The standard approach usually provides the anisotropy factor $\Delta$, so we incorporate this information in the MGD method and in this way, the use of the mimic constraint condition becomes unnecessary.

Using the analytical Tolman IV perfect fluid solution in the MGD approach we get a new solution that represents the anisotropic extension of Tolman IV solution. This new analytical solution satisfies all the usual criteria of physical acceptability. We have evaluated the physical consistency of our solution by examining the structure of matter sector, energy conditions, causality, the adiabatic index and stability against gravitational cracking. Therefore, this could be used to model actual stellar compact structures, such as neutron stars.

As a continuation of the study presented here, we suggest to explore of anisotropic solutions using the conservation equation obtained from the decoupling sector. This and other aspects will be considered in future works.

\end{document}